\documentclass[twocolumn,aps,pre,superscriptaddress,floatfix,nofootinbib]{revtex4-1}
\usepackage{graphicx}
\usepackage{epsfig}
\usepackage{thumbpdf}
\usepackage{amsfonts}
\usepackage{times}     
\usepackage{float}
\usepackage{indentfirst}   %
\usepackage{amsmath}
\usepackage{epstopdf}
\usepackage{textcomp}
\usepackage{tocvsec2}
\usepackage{epsfig}
\usepackage{xcolor}
\usepackage{comment}

\newcommand{\av}[1]{\left\langle {#1} \right\rangle}
\newcommand{\be}{\begin{equation}}
\newcommand{\ee}{\end{equation}}

\begin{document}

\title{Rapid decay in the relative efficiency of quarantine to halt epidemics in networks}

\author{Giovanni Strona} \email{goblinshrimp@gmail.com}
\affiliation{European Commission, Joint Research Centre, Directorate D - Sustainable Resources, Bio-Economy Unit, Via Enrico Fermi 2749, 21027 Ispra, Italy.}

\author{Claudio Castellano} \email{claudio.castellano@roma1.infn.it}
\affiliation{Istituto dei Sistemi Complessi (ISC-CNR), Via dei Taurini 19, 00185
 Rome, Italy}

\begin{abstract}
Several recent studies have tackled the issue of optimal network
immunization, by providing efficient criteria to identify key nodes to
be removed in order to break apart a network, thus preventing the 
occurrence of extensive epidemic outbreaks. 
Yet, although the efficiency of those
criteria has been demonstrated also in empirical networks, preventive
immunization is rarely applied to real world scenarios, where the
usual approach is the a posteriori attempt to contain epidemic
outbreaks using quarantine measures. 
Here we compare the efficiency of prevention
with that of quarantine in terms of the tradeoff between the number of
removed and saved nodes on both synthetic and empirical topologies.
We show how, consistent with common sense, but contrary to
common practice, in many cases preventing is better than curing:
depending on network structure, rescuing an infected network by quarantine could become inefficient 
soon after the first infection.
\end{abstract}

\maketitle

\section{Introduction}
\label{sec:introduction}

Networks provide a convenient framework for investigating epidemics, 
with nodes
corresponding to potential hosts of the infectious agent, and links
between nodes indicating potential routes for the spread of the
infection~\cite{PastorSatorras2015}. Much effort has been spent on
elaborating efficient procedures to immunize a
network~\cite{PastorSatorras2002, Cohen2003,Chen2008}.  A novel and
effective approach has been recently proposed~\cite{Morone15}, based
on the concept of optimal percolation. The idea is that, in order to
confine the infection and avoid large outbreaks, one must delete, in
the contact network, the minimal number of nodes sufficient to break 
the system into small, isolated fragments, with no giant component.

Although interesting from a theoretical perspective, the idea of
preventively reducing the risk of epidemics by
removing nodes from a network can sometimes be impractical. For
example, the modification of a transportation network~\cite{Tatem2006}
has deep economic and societal implications, apart from the
obvious technical difficulties.
In other scenarios, ethical aspects add
to the implementation challenges. For example, it would be very
controversial to designate specific individuals in the population for
mandatory vaccination. Similarly, in a spatial
network~\cite{Barthelemy2011} that maps infection pathways for
agricultural pests, it would be very difficult to require costly
preventive actions from some farmers only, especially in the case of a
purely hypothetical threat~\cite{Strona2017}.

In addition, different kinds of threat would modify the underlying
network, hence calling for different immunization strategies, not only
in terms of how to immunize a node, but also in terms of the identity
of nodes to be immunized. Considering the last example, a network of
infection pathways for an agricultural pest can be generated by
connecting all pairs of crops within a certain threshold distance,
representing how far a pathogen can spread from one crop to another
in a single step~\cite{Strona2017}. This distance, however, depends on
several aspects related to the pathogen's features (such as the flight
distance of its vectors), so that the same crops in a given region may
be arranged in a particular network topology for a certain disease, and 
in a very different one for another. In turn, this would lead to the
identification of different sets of critical nodes.

The discovery of a serious disease in the first phases of its spread
often leaves no other choice than to sacrifice individual
interests in the hope of preventing an outbreak with a coordinated
action. From a theoretical perspective, the simple removal of infected
nodes is a straightforward, trivial solution to the problem of
stopping an epidemic. However, in real-world scenarios, due to
infection latency (the interval between the time a host
becomes infectious, and the time it shows signs of disease), it is
common practice to try to isolate the infected nodes from the rest of
the network by also taking action on the surrounding nodes. For
example, in the case of the spread of disease in plants, it is usual
procedure to eliminate all the susceptible hosts close to an infected
one~\cite{Strona2017, Rodrigues2008}.  
Similarly, in the case of human diseases, a
typical intervention is the quarantine of asymptomatic
individuals that have had contacts with infected nodes (see, for
example,~\cite{Lipsitch2003}).

Some recent studies have investigated various aspects related to the
efficiency of different quarantine strategies. Nian and
Wang~\cite{Nian2010} considered ``high-risk'' immunization
(more properly, quarantine) of susceptible neighbors of infected individuals,
occurring with a probability $\delta$, and found that increasing
$\delta$ reduces the epidemic threshold of the 
Susceptible-Infected-Removed-Susceptible (SIRS) dynamics. 
Hasegawa and Nemoto~\cite{Hasegawa2017} studied numerically and 
analytically a similar strategy for Susceptible-Infected-Removed (SIR) 
dynamics, finding it more 
efficient at suppressing epidemics than random or acquaintance immunization.
Pereira and Young~\cite{Pereira2015} investigated, for 
Susceptible-Infected-Susceptible (SIS) dynamics,
the effect of the time period between the infection of an individual
and its isolation.

Apart from the difficulties that may arise in the practical
implementation of the two approaches (i.e. preventive immunization
vs. quarantine), a crucial question to be answered is: under which
circumstances is a preventive immunization strategy preferable to a
post-outbreak (quarantine) intervention?
To avoid confusion (the same terms are often used with different meanings
in the literature), let us specify that with quarantine
we intend the post-outbreak removal of
nodes which have been in contact with infectious individuals.
Immunization is instead the pre-outbreak removal of nodes based on some
a priori prevention strategy.

In this paper, we analyze the question by comparing the fraction of nodes 
saved using preventive immunization against the fraction of nodes that 
can be saved if quarantine is implemented at different times 
after the beginning of the outbreak.
We mainly focus on the conservative concept of
``complete" quarantine, i.e., we assume that the quarantine strategy
is implemented by removing all susceptible neighbors of infected nodes.
Moreover we assume that outbreak detection is immediate and perfect:
we always know which nodes are infectious. In reality early outbreak 
detection is a nontrivial problem for which clever strategies can
be devised~\cite{Eubank2004}.

We assess whether and when preventing an outbreak by 
immunization is a preferable strategy (in terms of saving more individuals) 
to attempting to rescue an infected network by quarantine. 
We consider a very simple theoretical framework (in terms of
epidemic dynamics, network topologies and assumptions
about the cost of immunization, quarantine, or being hit by the infection)
with the aim of grasping the main features of the problem. 
In contrast to previous approaches, the preventive immunization strategy
considered here is based on the recently introduced concept of
optimal percolation.

The overall resulting message of our study is that quarantine may be a
viable alternative only if enacted immediately after the start of the
outbreak. If the implementation is not prompt enough, then quarantine
becomes highly inefficient, as it stops the epidemic at the price of
removing many more nodes than those that would have been removed by
preventive immunization. In some cases it may even be more convenient
to let the disease propagate freely rather than applying a quarantine
measure. These conclusions hold very strongly for networks possessing
the small-world property (i.e. having small diameter). They hold
in a weaker form for networks
with large diameter (such as planar graphs): in that case the number 
of infected nodes grows more slowly and, as a consequence, quarantine 
remains convenient for longer times.

The ``complete" quarantine makes sense in many real world scenarios where
the topology of the contact pattern is not
known. This applies, for example, to the case of quarantining people
who have come into contact with a dangerous pathogen: although they
clearly belong to some sort of social network, its structure
would be, in most cases, elusive. 
Nevertheless, when the contact pattern is known, it is possible to
devise a more refined quarantine strategy, which removes a node
only if this saves on average more than one other node.
We investigate this aspect by introducing
an improved quarantine strategy that takes into account the average
effect, on the epidemic process, of whether or not a node is quarantined.
We show that the set of nodes that should be quarantined to halt an 
epidemic can be significantly smaller than the set of healthy neighbors
surrounding infected nodes. Nevertheless, even this improved quarantine 
strategy is not more efficient than preventive immunization.

The structure of the paper is the following. In Section II we define
the epidemic dynamics considered, the synthetic topologies of the underlying
interaction pattern, and give details on the
different approaches to be compared. We then compare the quarantine
strategy with two alternative courses of action: no intervention
(Section III) and preventive immunization (Section IV). Section V 
reports the same type of analysis for two real-world topologies.
The next Section is devoted to the presentation and performance 
evaluation of the improved quarantine strategy.
The last Section contains some concluding remarks, and an outlook
about future work.

\section{Epidemic model and countermeasures}
\label{sec:model}

We consider the Independent Cascade Model,
a parallel implementation of the SIR dynamics, the simplest model 
for the spreading of a pathogen conferring permanent 
immunity~\cite{Chenbook2013}.
Each individual sits on a node of a network and can be in one of 
three states: susceptible (S),  infected (I) or removed (R).
The total number of individuals is $N$.
At a given time step, a list of infected nodes is created and each node
in the list tries to infect each of his/her susceptible neighbors: 
each attempt is successful with probability $p$. 
After all nodes in the list have been considered, they all recover
(i.e. with probability $1$) switching to state R. Time is increased
by 1 and the next iteration begins with the compilation of the list 
of newly infected nodes: $p$ is the only parameter of the dynamics.
In the initial configuration a single randomly selected node is infected,
all others being susceptible. In a finite system the dynamics unavoidably
reach a final absorbing configuration with no infective nodes: all
individuals are either untouched by the epidemic (in state S) or
are recovered (state R) after having been infected.
The fraction $R$ of recovered nodes in the final state plays the 
role of the order parameter: for small values of $p$
the epidemic reaches only a small number of individuals around the
initial seed ($R$ is close to 0 for large systems); 
for large $p$ instead an extensive set of individuals is reached
by the infection, so that $R$ is finite in large systems.
The two regimes are separated by a critical value (the epidemic
threshold $p_c$) distinguishing for $N \to \infty $ between extensive 
spreading $R(p) >0$ for $p>p_c$ and small-scale local spreading 
$R(p) \to 0$ for $p \le p_c$.
We consider two classes of synthetic networks representing the static 
contact pattern among individuals. On the one hand there are
low-diameter networks (i.e. having the small-world property),
represented by Erd\"os-R\'enyi graphs or by scale-free networks
with degree distribution $P(k) \sim k^{-\gamma}$ with $\gamma=3$ and
$k_{min}=3$,
built using the uncorrelated configuration model~\cite{Catanzaro2005}.
The network size is in all cases $N=10^5$. The average degree is, unless
specified otherwise, $\av{k}=5$ ($\av{k}\approx 5.01$ for the
scale-free network).
The epidemic threshold is, with very good approximation, 
$p_c=\av{k}/[\av{k^2}-\av{k}]$~\cite{Dorogovtsev2008}.
On the other hand we also consider another type of network, which
does not possess the small-world property: a random
geometric graph, built in the two-dimensional square of unit size 
by connecting nodes at euclidean distance smaller 
than $d=0.00395$ ($\av{k}\approx 5.04$).
In this topology the diameter grows asymptotically as a power of the
number of nodes~\cite{Ellis2007}, as in a regular lattice.

Most of our work regards the comparison between a preventive
immunization protocol and a quarantine strategy implemented only after
the outbreak has started.

The preventive immunization is based on the concept of optimal
percolation proposed by Morone and 
Makse~\cite{Morone15}\footnote{Notice that, despite the formulation
of the original paper, optimal percolation is a strategy for minimizing
(not for maximizing) spreading in a network~\cite{Radicchi2017}.}. 
Extensive epidemic outbreaks in a network are possible only if a giant 
component (i.e. a cluster of connected nodes covering a finite fraction of the
structure) is present. Otherwise, only small outbreaks can occur,
even for extremely infective pathogens ($p=1$).
Therefore, by removing a number of nodes sufficient to dismantle
the giant component it is
possible to prevent large outbreaks. By means of different 
strategies~\cite{Morone15, Braunstein16, Clusella16, Zdeborova16}
it is possible to reduce this number to values close to the theoretical
minimum, thus providing an optimal preventive immunization.
If the removal of a fraction $n_r$ of nodes destroys the
giant component one can consider that a fraction $n_{OP}=1-n_r$
of nodes is effectively {\em saved} from infection.
In Appendix A we provide details on how $n_{OP}$ is
determined in our study.

The alternative strategy to contain the epidemic and save individuals
is quarantine, which is intended here as the removal at a given time step
of all susceptible neighbors of all infected nodes. 
This procedure leads to the immediate end of the epidemic, 
leaving a fraction $n_Q$ of the nodes saved, i.e. not reached by
the contagion. In practice, it is possible to calculate the 
fraction $n_Q$ for each time step as the epidemic unfolds by simply
counting the number of susceptible nodes which are not direct neighbors
of infected nodes.

In the following we compare the fraction of nodes saved by quarantine
with the fraction of those saved by optimal percolation or not
reached by the epidemic in the absence of any intervention.
The underlying assumption is that immunizing a node or quarantining it 
has the same cost of the node being infected. 
This might be an appropriate scenario for, say, an incurable deadly tree 
disease, for which immunization or quarantine both imply the physical 
felling of the tree. 
In other cases where the cost of the different actions is not the same,
the analysis is more complicated, and beyond the scope of this paper.

\section{Comparison between quarantine and no intervention}

The first question we address is whether a quarantine strategy
is any better than doing nothing. The answer to this question
is not completely trivial: for small values of the spreading
probability $p$, the epidemic soon becomes extinct spontaneously, touching
a very limited number of individuals; a quarantine intervention 
in such a case implies the removal of more individuals than those
likely to be reached by the epidemic. 
For larger values of $p$ it's a matter of timeliness: if the
intervention is quick enough, quarantine may be useful; otherwise, 
it is more costly than leaving the epidemic spontaneously evolve and 
disappear.

\begin{figure*}
\includegraphics[width=\textwidth]{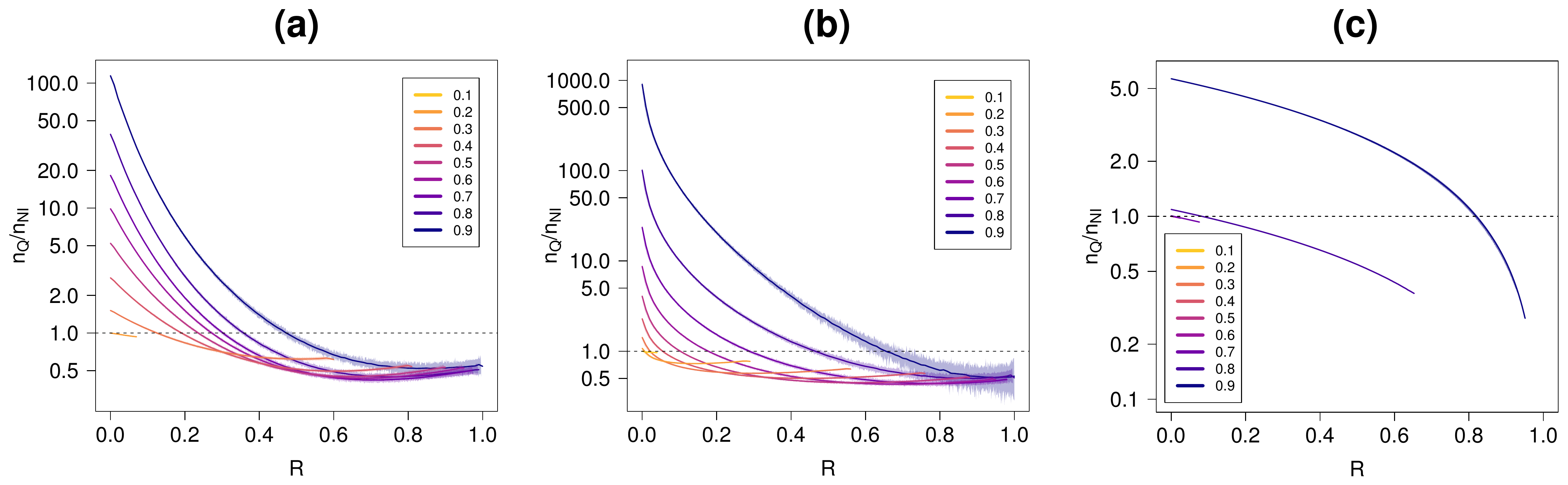}
\caption{(color online) (a) Ratio between the fraction of nodes $n_Q$ saved by
  quarantine and the fraction of nodes $n_{NI}$ not reached by
  the infection in the absence of any intervention as a function of
  the fraction $R$ of nodes recovered at the time the quarantine is
  implemented. The network is an Erd\"os-R\'enyi graph with $N=10^5$ nodes. 
  The curves are for growing 
  values (bottom to top) of the spreading probability $p$. 
  The dashed line indicates when the ratio is 1, i.e. the value discriminating
between convenient quarantine (ratio $>1$) and inconvenient
quarantine (ratio $<1$). 
  (b)
  The same plot for a scale-free network with $\gamma=3$.
  (c)
  The same plot for a random geometric graph.
  For this figure (and in all the others) we performed 10000 realizations of
  the process. Solid lines are average values computed over 100 intervals
  (of width $0.01$), 
  while the shaded areas around them are the variations between the minimum 
  and the maximum in 1000 intervals.}
\label{Fig1}
\end{figure*}

This is confirmed in Fig.~\ref{Fig1}, where we plot the ratio between 
the fraction $n_Q$ of nodes saved by the quarantine strategy and the 
fraction $n_{NI} = 1-R(t \to \infty)$  of nodes which are not reached by the 
infection in the absence of any intervention. Unsurprisingly, it turns 
out that for $p$ up to the epidemic threshold quarantine is never 
convenient: it removes more individuals than those that would be hit
by the freely evolving epidemics.
For larger values of $p$, quarantining may be very useful, but only if 
promptly implemented:  for example, for an Erd\"os-R\'enyi graph and $p=0.3$,
the implementation of quarantine  actually saves more individuals 
than those that would have survived with no-intervention, 
only if applied as long as $R<0.12$. 
For a scale-free network, the effect is even more noticeable.
A different scenario applies instead for the random geometric
graph.
In that case, above the threshold (which is of the order of 0.8). 
the ratio $n_Q/n_{NI}$ remains very large up 
to values of $R$ close to 1: quarantine is more convenient than
no intervention practically at any time during the epidemic.

\section{Comparison between quarantine and preventive immunization}
We now evaluate the efficiency of a quarantine procedure
compared to that of a preventive immunization strategy based on optimal
percolation. In practice, the latter is implemented by removing a 
fraction $1-n_{OP}$ of the nodes,
breaking the network into small clusters so that any outbreak cannot
grow large, (see Appendix A for details).
In particular, we compare the global fraction $n_{OP}$ of nodes
saved by applying the preventive immunization strategy with the
fraction $n_Q$ saved by applying the quarantine strategy. Also
in this case, the efficiency of quarantine depends on its timeliness,
i.e. on the value of $R$ at the time it is applied.
\begin{figure*}
\includegraphics[width=\textwidth]{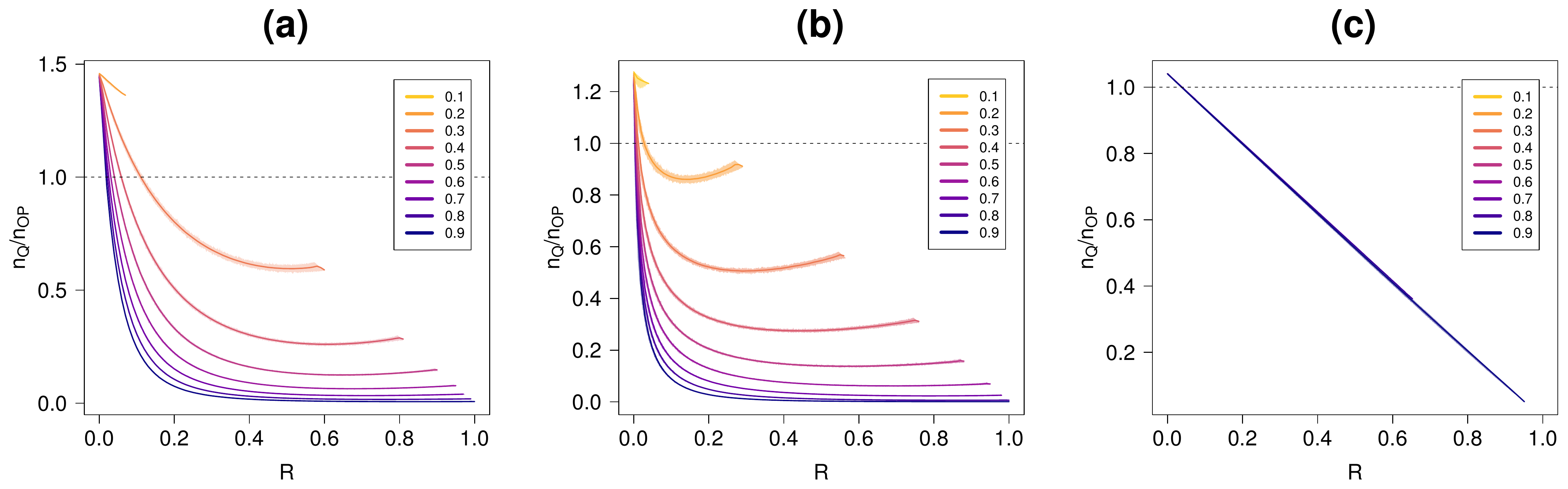}
\caption{(color online) (a)
Ratio between the fraction $n_Q$ of nodes 
saved by quarantine and the fraction of nodes $n_{OP}$ saved by preventive 
immunization, as a function of the fraction $R$ of
nodes in state R when quarantine is applied. 
The network is an Erd\"os-R\'enyi graph with $N=10^5$ nodes.
The curves are for growing values (top to bottom) of the spreading 
probability $p$.
The dashed line indicates when the ratio is 1, i.e. the value
discriminating between convenient quarantine (ratio $>1$) and 
inconvenient quarantine (ratio $<1$). 
(b) The same plot for a scale-free network with $\gamma=3$.
(c) The same plot for a random geometric graph.
}
\label{Fig2}
\end{figure*}
Fig.~\ref{Fig2} shows that, for the two small-world networks, 
$n_Q$ decays very rapidly below $n_{OP}$: unless applied in the very 
early stages of the epidemic, quarantine is less efficient than 
preventive immunization.
The only exception to this rule occurs for small values of $p$:
in such cases the epidemic spontaneously becomes extinct after relatively
short times, so that preventive immunization is never more efficient.
Again, the effect is more dramatic when the topology is scale-free:
quarantine becomes highly inconvenient as soon as only a few percent of
the population have been hit by the epidemics.
For the random geometric graph, the effect is once again
considerably milder.
The ratio decays linearly with $R$ with the same slope for any value 
of $p$. Hence the initial window for an efficient quarantine intervention
has the same width independent from $p$.

A qualitative understanding of the phenomenology is straightforward.
Starting from the initial seed the epidemics reaches at time $t$
nodes at distance $t-1$ from the seed. In a network with small-world
property the number of these nodes grows exponentially, approximately as 
$[p(\av{k^2}/\av{k}-1)]^{t-1}$. 
In few steps the whole population is reached and hence the application
of quarantine is highly ineffective, apart from the very first moments
of the spreading process.
In a planar network instead the same quantity
grows much more slowly (quadratically) in time. Hence quarantine
remains effective up to considerably longer times.

It is possible to analytically estimate, in both scenarios,
the value $R^*$ after which the quarantine strategy becomes less 
efficient than preventive immunization. 

For the random regular network, given the quasi planar structure,
the set of removed nodes forms approximately a circle around the seed, 
its radius growing over time. Infected nodes form a ring sitting at
the boundary of the circle and nodes to be quarantined
another ring, surrounding the first. For this reason
the fraction $I$ of infected nodes is proportional to 
$\sqrt{R}$ and the same the fraction $Q$ of nodes to be quarantined.
As soon as the number of removed nodes is of the order of a few dozens, 
$I$ and $Q$ can be neglected with respect to $R$.
Hence the fraction of saved nodes $n_Q$ is
\begin{equation}
n_Q = 1-R-I-Q \approx 1-R
\end{equation}
This formula fully explains the linear behavior and the lack of 
dependence on $p$ observed in Fig.~\ref{Fig2}(c).
The values after which quarantine is inefficient is $R^*=1-n_{OP}$.

\begin{figure}
\includegraphics[width=.98\columnwidth]{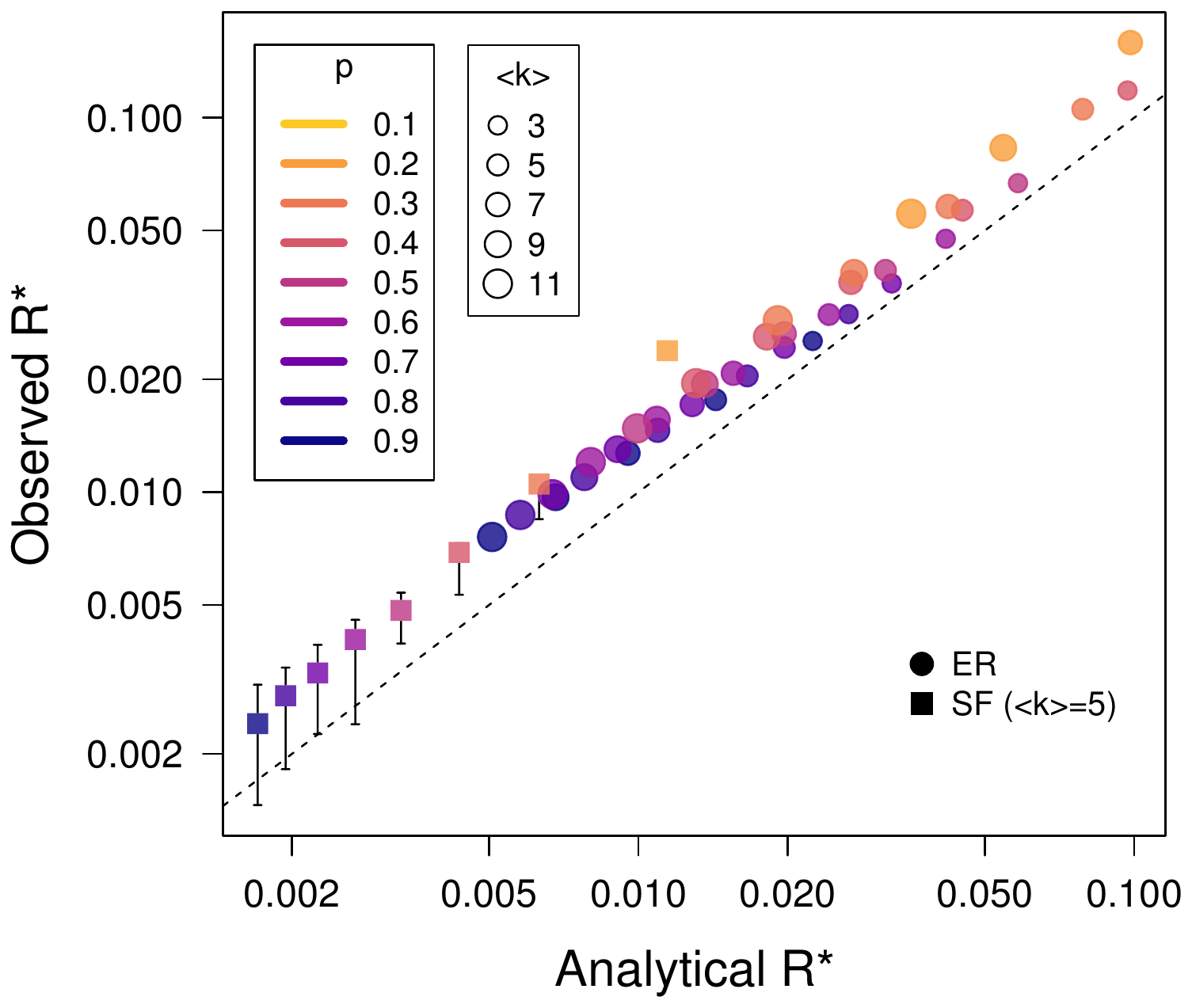}
\caption{Comparison between analytical predictions (Eq.~(\ref{Rstar})) 
and simulation results for the fraction $R^*$ of recovered nodes
after which quarantine becomes less convenient than preventive
immunization. The dashed line indicates perfect equality. 
Simulations are performed on Erd\"os-R\'enyi (ER) networks with
various values of the average degree $\av{k}$ and on a
scale-free (SF) network. 
Error bars are not shown when smaller than symbols size.
System size is $N=10^5$.}
\label{Fig3}
\end{figure}

The derivation of the same relation for locally small-world 
(locally tree-like) networks 
is more involved. A calculation based on the heterogeneous mean-field 
(HMF) approximation for SIR dynamics, is presented in Appendix B, yielding:
\be
n_Q(R) = 1 - \left( 1+ \frac{1}{\tau} \frac{\av{k^2}}{\av{k}} \right) R,
\label{nQ}
\ee
where
\be
\tau=\frac{1}{p \av{k^2}/\av{k}-(p+1)}.
\ee
By equating this expression to $n_{OP}$ we obtain
\be
R^* = \frac{1-n_{OP}}{1+\frac{1}{\tau} \frac{\av{k^2}}{\av{k}}}.
\label{Rstar}
\ee
In Fig.~\ref{Fig3}, we compare this analytical prediction with
the results obtained from numerical simulations on both
ER and scale-free networks. The agreement is satisfactory, 
also considering the known discrepancies between 
HMF theory for SIR and simulation 
results~\cite{PastorSatorras2015,Castellano16}.
We note that the value of $R^*$ is 
reduced as the average degree of homogeneous networks is increased
or as the network becomes more heterogeneous.

\begin{figure}
\includegraphics[width=.98\columnwidth]{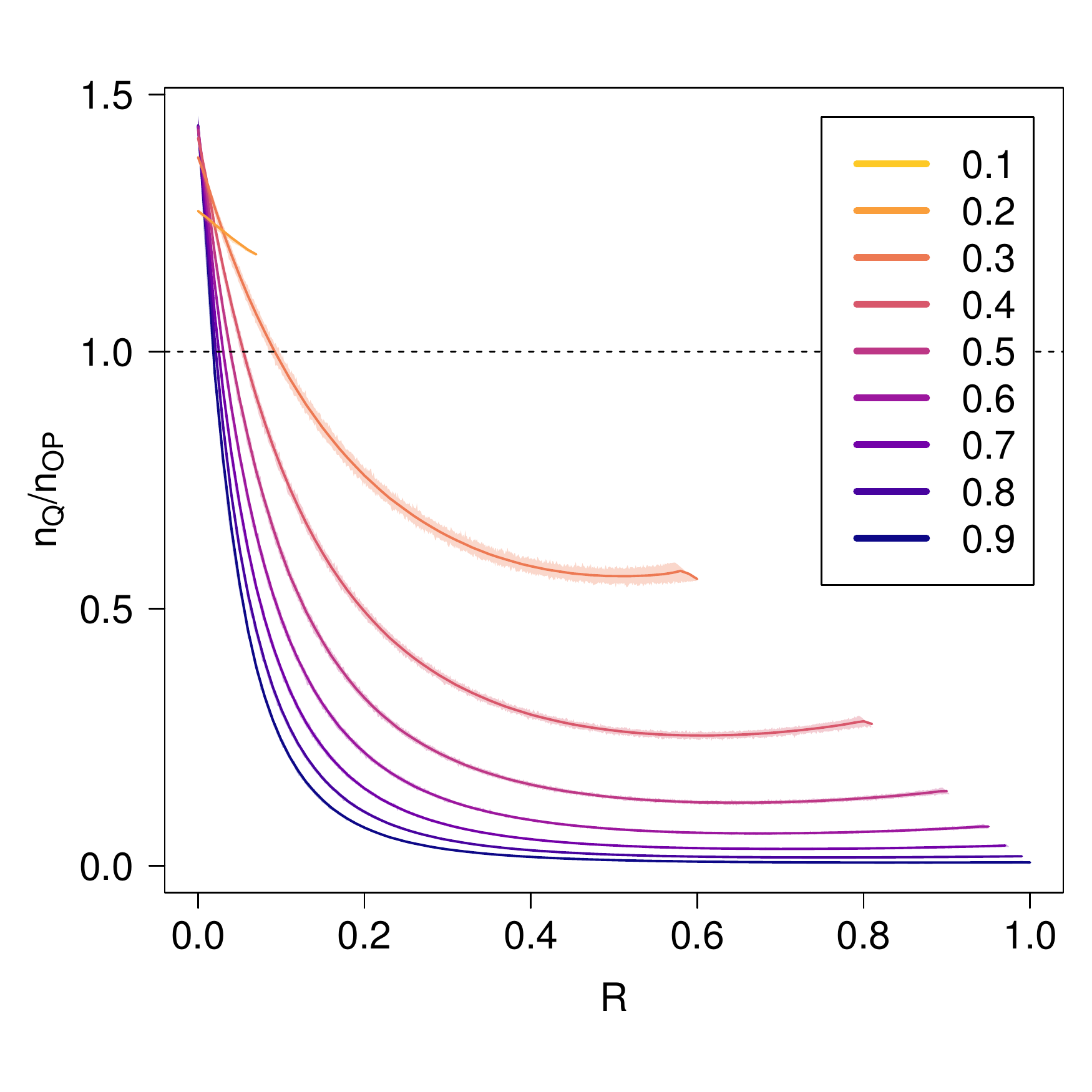}
\caption{(color online) Ratio between the fraction $n_Q$ of nodes saved by 
quarantine and the fraction of nodes $n_{OP}$ saved by ``informed" preventive 
immunization (i.e. with $p$ known a priori), as a function of the fraction of
nodes removed when quarantine is applied. 
The network is an Erd\"os-R\'enyi graph with $N=10^5$ nodes.
The curves are for growing values (top to bottom) of the spreading
probability $p$.}
\label{Fig4}
\end{figure}

In Fig.~\ref{Fig2} preventive immunization is implemented by optimal
percolation, i.e., by removing the minimum number of nodes sufficient
to destroy the giant component of the network, thus precluding the
possibility of endemic outbreaks, even in the extreme hypothesis that
the transmission probability is $p=1$.  If one has information about
the actual value of $p$ it is possible to use it in the preventive
immunization strategy.  In that case, one can compare $n_Q$ with the
average fraction of nodes saved, given that the spreading probability
is $p$. Clearly $n(p) \ge n_{OP}$ [with $n(p=1)=n_{OP}$].  See
Appendix A for details about the calculation of $n(p)$.  The ratio of
the fractions of saved nodes with the two strategies (``informed"
preventive immunization vs quarantine) is displayed in
Fig.~\ref{Fig4}.
In this scenario, where $p$ is known a priori, the observed patterns
are very similar to those observed assuming $p=1$; the higher 
convenience of preventive immunization is just slightly more pronounced. 
Moreover, one must consider that $n(p)$ is only the {\em
average} fraction of saved individuals. Depending on the specific
realization of the epidemic process, the actual value can be
larger but also smaller.  Only the immunization assuming
$p=1$ is guaranteed to save a fraction $n_{OP}$ of the nodes. Since
the patterns observed assuming $p=1$ (Fig.~\ref{Fig2}) are very
similar to those observed for $p$ known a priori (Fig.~\ref{Fig4}),
i.e. the ``informed" (and hence partial) immunization does not, 
on average, lead to the saving of a much larger number nodes 
than those saved by the
optimal percolation, the latter should be, in general, preferred.

\section{On real-world topologies}
In this Section we check whether the same phenomenology observed
in synthetic networks also occurs in real-world topologies, where
correlations, clustering, mesoscopic structures add complexity to the picture.
We consider an instance of a small-world network with small diameter,
an undirected version of the World airline Network (WAN)~\cite{wan}. 
The number of nodes is $N=3257$, the average degree $\av{k}\approx 11.76$
and the diameter is $12$.
We also consider the US power-grid network~\cite{Watts1998},
 a quasi-planar graph 
with large diameter ($46$), $N=4941$ nodes and $\av{k}\approx 2.7$.

\begin{figure*}[!htb]
\includegraphics[width=\textwidth]{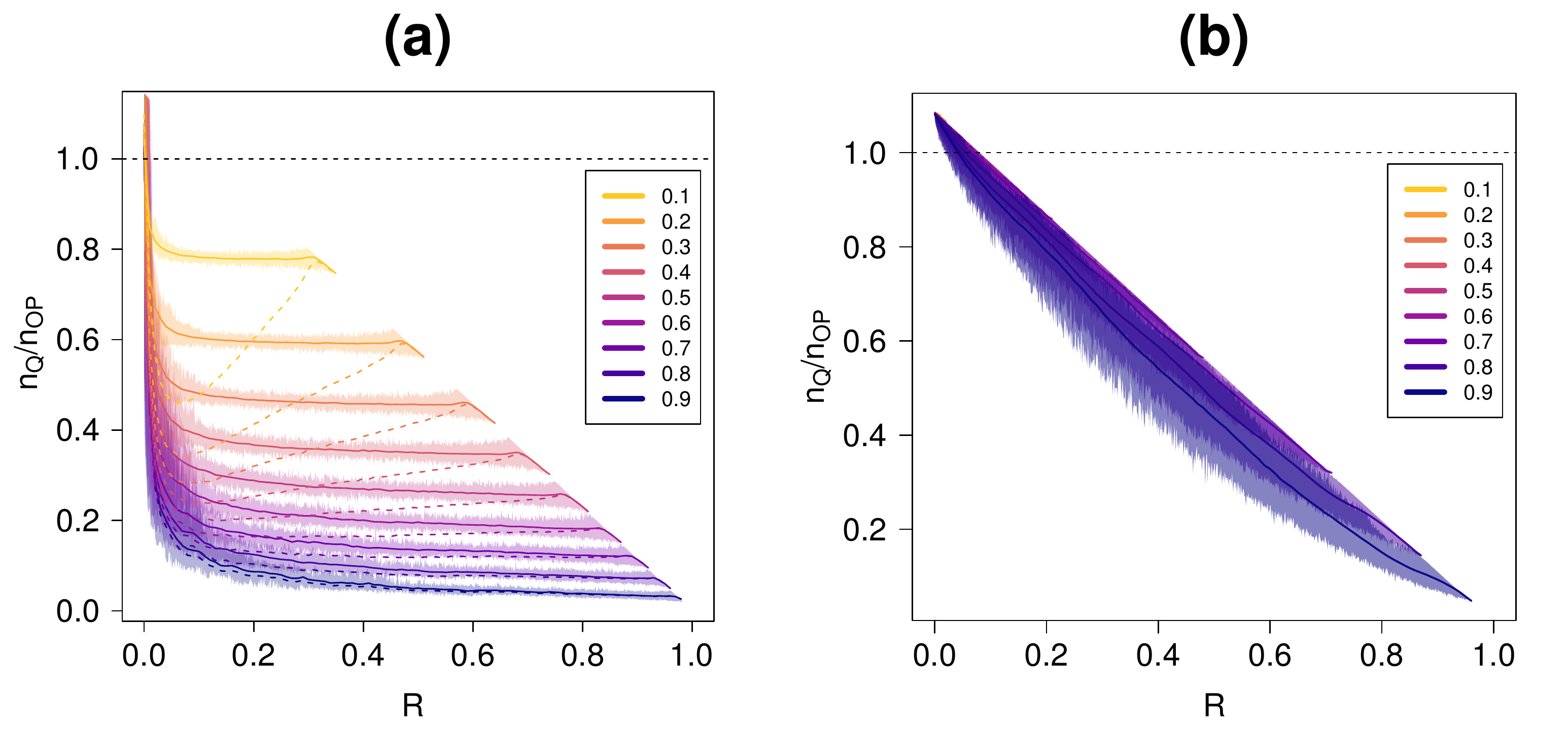}
\caption{(color online) Ratio between the fraction $n_Q$ of
nodes saved by quarantine and the fraction of nodes $n_{OP}$ saved
by preventive immunization, as a function of the fraction of
nodes removed. The topology is (a) WAN; (b) US power-grid. In the
WAN topology (a)
we report both the results of the simulations implementing either
the complete quarantine (dashed lines), or the improved one (solid
lines).
The curves are for growing values (top to bottom) of the spreading
probability $p$.}
\label{fig6}
\end{figure*}

We show the comparison between the quarantine strategy and preventive 
immunization in Figs.~\ref{fig6}.

The figure confirms the same qualitative behavior observed in synthetic
networks. For small-world networks quarantine becomes,
almost immediately after the beginning of the outbreak,
a less efficient strategy than preventive immunization. 
For all the considered values of $p$, the observed $R^*$ values are
$\approx 0.001$ (i.e. $\approx 3$ nodes),
close to the analytical estimates ($\approx 0$ nodes).
In the case of the quasi-planar network instead the dependence
of $n_Q$ on $R$ is almost linear and as a consequence there is
a finite (yet small) range of $R$ values for which quarantine is
more efficient than preventive immunization.

Also the comparison of quarantine with non intervention (not shown)
confirms the qualitative behavior observed in synthetic networks.

\section{Improved quarantine strategy}

In the previous Sections we have seen that a post-outbreak quarantine
strategy is almost always less convenient (in terms of saved individuals) 
than optimal preventive
immunization. So far, however, we have considered a crude quarantine
strategy, consisting in the removal of {\em all} susceptible neighbors
of infected individuals. This approach makes sense in many real world
situations, where the complete topology of the network
is not known a priori.  For example, it is
reasonable to quarantine all people that have been exposed to a
dangerous pathogen because, in most cases, the structure of the social
network they belong to is not well known.  However, in other situations
where the topology is available, the complete quarantine may not
be the best strategy compared to a more parsimonious approach aimed at
avoiding the removal of a particular node unless it brings benefit to
other nodes. A trivial example of such a situation is that of a
healthy node completely surrounded by infected neighbors. Removing the
healthy node from the network will not affect the probability of
infection of other uninfected nodes.  Thus, rather than
removing it from the network, it would be more
reasonable to leave it untouched, hoping it remains uninfected.

General criteria for the identification of nodes that are not to be
quarantined may, however, be less self-evident. We consider a simple
one, based on the evaluation of $I_q(V_q)$, i.e. the number of nodes
that will be, on average, reached by the infection if
the candidate $V_q$ for quarantine is left untouched. 
If this number is smaller than 1, quarantining $V_q$ would lead
to a {\em reduction} in the number of saved individuals: hence 
it is more convenient not to quarantine it.

For the independent cascade model with transmission probability 
$p$ on a tree structure, the average number of nodes that
will be infected if a node $V_q$ candidate to be quarantined is
not removed is
\be 
I_q(V_q)= p \sum_{l=1}^z p^l u_l.
\label{Iq}
\ee
Here $p$ is the probability that $V_q$ actually becomes infected,
$u_l$ is the number of nodes at exactly $l$ steps from $V_q$, 
and $z$ is the path length from $V_q$ to the farthest
uninfected node.
If $I_q>1$, then $V_q$ should be quarantined.  Otherwise, leaving $V_q$
untouched would result, on average, in less than one node being infected. 
Hence, a higher number of saved nodes is obtained if $V_q$ is not quarantined.
On a network that is not a tree, Eq.~(\ref{Iq}) is no longer exact.
However, one can use it as an approximation, provided that
$u_l$ is computed after the exclusion from the network of all other
nodes candidate for quarantine, as well as of all removed ($R$) nodes. 
Moreover, we can replace the first $p$ factor with the probability
$p_q=[1-(1-p)^{n_q}]$ that the infection is transmitted to $V_q$ by one
of its $n_q$ infected neighbors.

As in the case of the ``informed" preventive immunization, 
$I_q$ with $p<1$ will provide information on the average expected effect 
of not quarantining a node, i.e. it
will not exclude the possibility that a much larger number of nodes 
than the expected value is actually reached by the infection.
Conservatively setting $p=1$ allows for the identification of the trivial 
cases of $V_q$ nodes surrounded by infected nodes only.

Another limitation of this simple approach is that in non-tree networks
there will generally be an overlap among sets of nodes protected by 
different $V_q$ nodes. 
Especially for high values of $p$, there could be situations
where a high number of $V_q$ nodes are simultaneously protecting a
small number of healthy nodes. In such cases, $I_q(V_q)$ will likely be
greater than 1 for most of the candidates $V_q$, but quarantining
all of them will save a small number of nodes at the expense of removing
many.
Avoiding the quarantine of all the $V_q$ nodes that protect the same (small) 
set of healthy nodes could be the most efficient strategy. 
A very simple procedure for taking this into account
is implemented by excluding from the network all infected and removed nodes,
and then counting the expected number of $V_q$ nodes that will be infected
(i.e. the sum of their $p_q$ values) in any network component. If that
sum plus the number of healthy nodes in the component
is smaller than the total number of $V_q$ nodes in the component, no
$V_q$ nodes in the component should be quarantined.

\begin{figure}
	\includegraphics[width=.98\columnwidth]{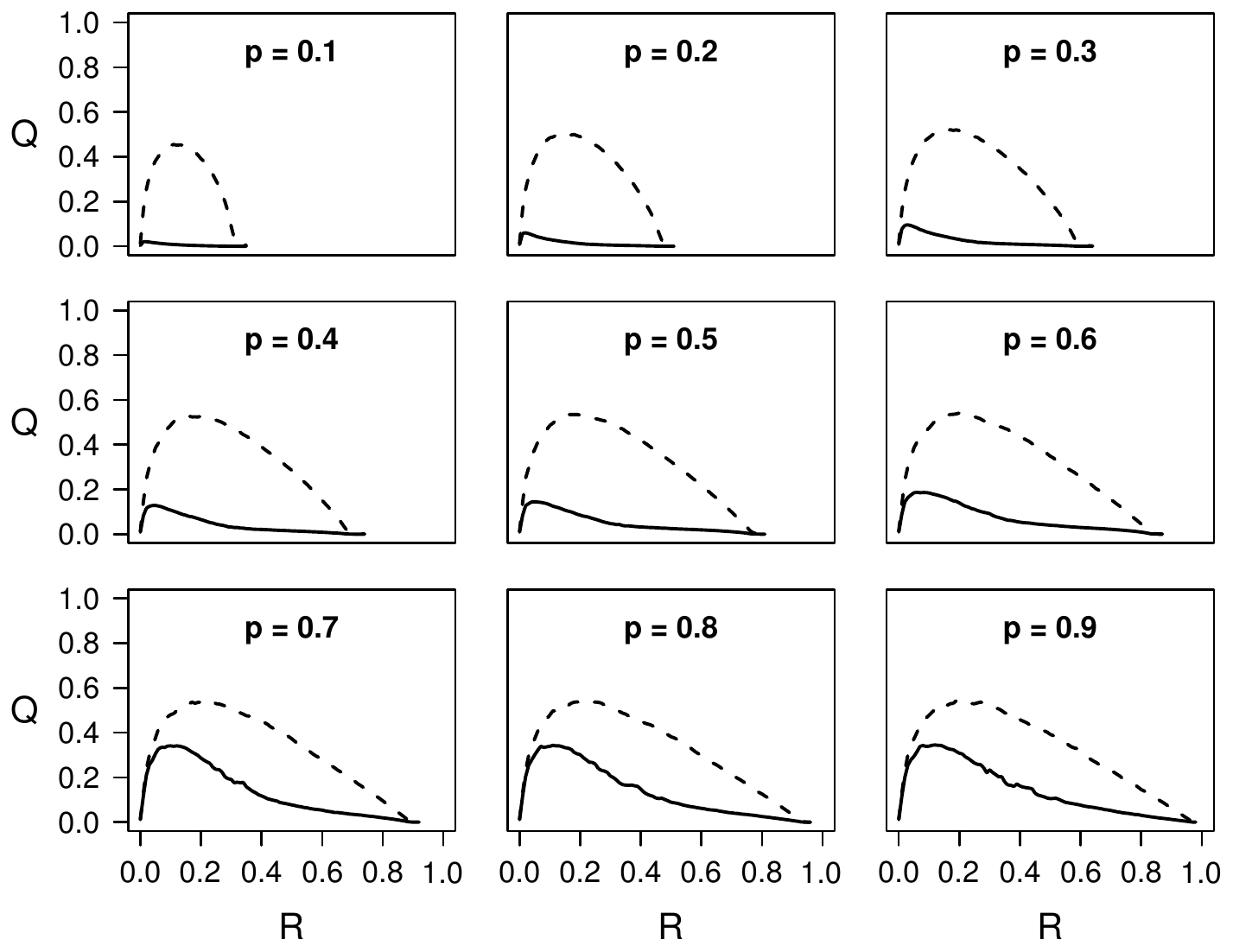}
	\caption{Comparison of the number of nodes to be quarantined
          $Q$ as a function of $R$ in the case of complete quarantine
          (dashed lines) or 
          improved quarantine (solid lines). Different panels refer to 
          different values of the spreading probability $p$.
          The topology is the WAN.}
	\label{fig7}
\end{figure}

We tested the combination of these criteria on the WAN topology.
In Fig.~\ref{fig7} we show how the method
substantially reduces the number of nodes to be quarantined.
In Fig.~\ref{fig6} we show how this, although it translates into a net
improvement in the average fraction of nodes saved by the quarantine,
does not affect our main conclusion: preventive immunization is still
much more efficient than quarantine shortly after the first infection.

\section{Conclusions}
\label{sec:conclusion}

In this paper we have compared the efficiency of quarantine protocols
with respect to optimal prevention strategies in terms of the number
of individuals that remain unaffected by the spread of the epidemic.
We find that almost always preventing is better than curing, unless
the quarantine approach is applied in the very early stages of the
spreading process. The time window for quarantine to be convenient
is extremely short for topologies with small diameter 
(possessing the small-world property)
and is further reduced if the average degree is increased or the 
degree distribution is broader.
If quarantine is applied too late it may even become more costly than
leaving the to contagion evolve freely until its spontaneous disappearance.
Quarantine performs better in networks with large diameter such as
planar graphs and the like. In such a case the number of infected
nodes does not grow exponentially with time and this makes the quarantine
strategy relatively efficient for longer intervals.

These results provide a basic understanding of the salient features
of the problem. 
A number of limitations must, however, be clearly spelled out.

First,
there are many epidemics where the infection can be transmitted from one
individual to another, even before symptoms permitting the
identification of infection show up. For example, in the case of
vector-borne diseases, a newly infected host may become able to
transmit the disease to a vector before the signs of the disease
become detectable. In such a situation, limiting the attention to the
nearest neighboring nodes could be not enough to ensure that the
spread is stopped.

Second, the identification of the optimal percolating set 
requires full knowledge of the contact network. If the topology is not
completely known it is impossible to immunize optimally; conversely,
quarantine is possible, as it requires only local information.
On the other hand, if the quarantine is applied too late, it might
be not only less efficient than prevention, but also too resource
demanding, as it requires the simultaneous treatment of an exponentially
large number of individuals.

Finally, it will be interesting to lift the assumption that the costs
of being infected, immunized or quarantined are the same. More realistic
scenarios, where these costs differ (and may even depend on
the infection stage), may give rise to a rich and interesting 
phenomenology.

\section*{Acknowledgments}
The views expressed are purely those of the writers and may not in any 
circumstance be regarded as stating an official position of the European Commission.

\section*{Appendix A: Immunization strategy based on optimal percolation}

The preventive immunization strategy is based on the concept
of optimal percolation, i.e. the identification
of the smallest possible set of nodes such that their removal splits
the whole topology into a set of mutually disconnected clusters of
nonextensive size.
If nodes belonging to this optimal set are removed, it is physically
impossible for any outbreak to reach a finite fraction of
the network, so no global epidemic is possible.
To implement this approach we use the procedure described by 
Morone and Makse~\cite{Morone15}. 
We remove sequentially one by one all nodes in the network based on
their ‘collective-influence’, which is recalculated at each step (using a
radius of 3, that is, considering in the computation all nodes within
three steps from the target node). 

At any  step $t$ we take note of the number of nodes still in the 
network ($N_t$) and of the size ($S_{q,t}$) of each connected network 
component ($q$). 
In the case of an epidemic starting from a random node 
in the network at step $t$, the probability that a component 
$q$ is infected is equal to $S_{q,t}/N_t$. 
If the seed belongs to $q$, the number of nodes saved will 
be equal to $N_t - S_{q,t}$. 
Thus, after removing $t$ nodes, the expected number $X_t$ of saved 
nodes is 
\be
X_t = \sum_{q} \frac{S_{q,t}}{N_t} \left(N_t - S_{q,t}\right) =
N_t \left(1-\frac{\sum_q S^2_{q,t}}{N_t^2} \right).
\ee
$X_t$ will in general be low at the beginning of the immunization, 
when the size of the largest connected component is still very large. 
Then, it will increase until it reaches a maximum, representative of 
the best tradeoff between the number of nodes removed, and the level of
immunization reached.  After that optimal point, additional node removals
will not lead to an overall reduction of network components’ size 
(so that $X_t$ will decrease approaching 0).

\begin{figure}
\includegraphics[width=.98\columnwidth]{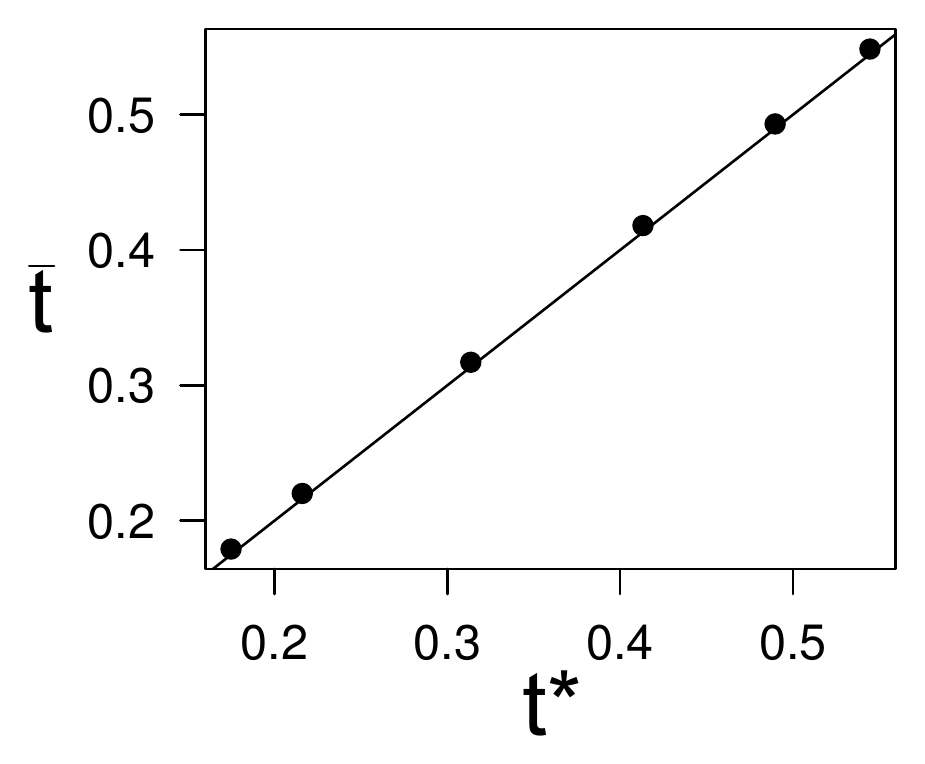}
\caption{Scatter plot of $t^*$ vs $\bar{t}$
for ER graphs with various $\av{k}$ and scale-free with $\gamma=3$
and $k_{min}=3$ ($\av{k} \approx 5.01$) showing that the two
quantities are almost identical. The solid line indicates perfect
equality.}
\label{fig8}
\end{figure}

We assume the number $t^*$ corresponding to the maximum value of $X_t$ 
as the number of nodes sacrificed by a preventive immunization approach. 
Fig.~\ref{fig8} shows that this value $t^*$ almost perfectly matches the 
value $\bar{t}$ calculated via the usual criterion, as the number
of nodes to be removed such that the largest connected cluster remaining
is smaller than $\sqrt{N}$~\cite{Clusella16}.

This indicates that $X_t$ plays the role of susceptibility for the optimal
percolation transition.

The procedure just described guarantees that, after removing $\bar{t}$
nodes, no outbreak can reach endemic proportions, even in the worst
case $p=1$. If the spreading probability is smaller than 1, however,
the fraction of saved nodes will be on average $n(p)$ larger than 
$n(p=1)=n_{OP}$. To determine this quantity the procedure is the following. 
At step $t$ (i.e. after $t$ nodes have been removed based 
on their collective influence), some SIR simulations are performed 
and the average number of saved nodes is computed. This quantity tends
to grow with $t$ due to the immunization, but for larger $t$ values it
decreases because of the node removal. The value of $t$ corresponding
to the maximum sets the optimal value $n(p)$ used in Fig.~\ref{Fig2}.

\section*{Appendix B: Analytical evaluation of $R^*$}
In this Appendix, we present the details of the derivation
of Eq.~(\ref{nQ}), based on the heterogeneous mean-field (HMF) 
theory for SIR~\cite{Boguna2003}. Alternative approaches
based on message-passing~\cite{Rogers2015} or branching 
processes~\cite{Gleeson2016} would be possible.

We define as $S_k$ (respectively, $I_k$, $R_k$) the density of nodes
of degree $k$ which are in state S (I, R, respectively) at time $t$.
The total fraction of nodes in state S is $S = \sum_k P(k) S_k$ 
(analogous formulas hold for the other states).
For simplicity, we omit the explicit indication of the time dependence,
where not needed.
For continuous dynamics 
[defined by the parameters $\mu$ 
(rate of spontaneous recovery) and $\beta$ (rate of infection through
an edge connecting a node in state I with a node in state S)] 
the evolution of the densities is, within the HMF approximation, given by
\begin{eqnarray}
\dot{S}_{k} &=& -\beta k S_{k} \Theta \\
\dot{I}_{k} &=& -\mu I_k + \beta k S_{k} \Theta \\
\dot{R}_{k} &=& \mu I_{k} \,,
\label{eq:hmf}
\end{eqnarray}
where
\be
\Theta = \frac{1}{\av{k}} \sum_k P(k) (k-1) I_k
\ee
is the probability that an edge points to an infected vertex and is 
capable of transmitting the disease.
By rescaling the time unit it is possible to set $\mu=1$ with
no loss of generality.
The form of the equations guarantees the conservation of the total 
probability $S_k+I_k+R_k=1$. The initial condition with a single infected
node implies $S_k \approx 1$, $I \approx 0$, $R_k=0$.

Multiplying the equation for $I_k$ by $P(k) (k-1)/\av{k}$ and summing
over $k$ we obtain
\be
\dot{\Theta} = -\Theta + \beta \Theta \frac{1}{\av{k}} \sum_k P(k) (k-1) k S_k.
\ee
For sufficiently short times one can assume $S_k \approx 1$, so that
\be
\dot{\Theta} = -\Theta + \beta \Theta 
\left( \frac{\av{k^2}}{\av{k}}-1 \right) = \frac{\Theta}{\tau}.
\label{eqTheta}
\ee

The temporal scale 
\be
\tau  = \frac{\av{k}}{\beta \av{k^2}-(\beta+1)\av{k}}
\label{tauapp}
\ee
governs the exponential growth (or decay) 
of $\Theta$, depending on whether $\beta$ is larger or smaller than
the threshold $\beta_c = \av{k}/(\av{k^2}-\av{k})$.

By integrating Eq.~(\ref{eqTheta}) and inserting the result
$\Theta(t)=\Theta(0) e^{t/\tau}$ into the equation for $I_k$ we get
\be
\dot{I}_k = -I_k + \beta k \Theta(0) e^{t/\tau},
\ee
which can be integrated yielding
\be
I_k(t) = I_k(0) e^{t/\tau},
\label{eq:I_k}
\ee
where $I_k(0) = k \Theta(0) \av{k}/(\av{k^2}-\av{k})$.

By integrating the equation for $R_k$ [Eq.~(\ref{eq:hmf})] we obtain
\be
R_k = \int_0^t dt' I_k(t') = I_k(0) \tau (e^{t/\tau}-1) \approx \tau I_k,
\ee
so $I_k = R_k/\tau$.

The quarantine procedure consists in removing from the system all infected
nodes and all their susceptible neighbors. For infected nodes of degree 
$k$ this  means removing $(k-1) I_k$ nodes; the factor $k-1$ accounts for 
the fact that one neighbor of an infected node is the node that transmitted
the infection to it, therefore it cannot be susceptible.
Hence the global fraction of nodes which are left after the quarantine
procedure is
\be
n_Q = \sum_k P(k) [1-R_k-I_k-(k-1) I_k] = 1 - R - \sum_k P(k) k I_k
\label{n_Q}
\ee
The last term on the right hand side in Eq.~(\ref{n_Q}) can be rewritten,
by considering the temporal evolution of $I_k$ [Eq.~(\ref{eq:I_k})], as
\be
\sum_k P(k) k I_k = I \frac{\av{k^2}}{\av{k}},
\ee
which, inserted into Eq.~(\ref{n_Q}) and using $R = \tau I$, yields
the expression for the total fraction of nodes $n_Q$ saved by a quarantine
applied when the number of recovered nodes is $R$:
\be
n_Q(R) = 1 - \left( 1+ \frac{1}{\tau} \frac{\av{k^2}}{\av{k}} \right) R .
\label{n_Q2}
\ee

It is important to note that this expression holds for
short times and on locally tree-like networks (otherwise the expression
$(k-1) I_k$ for the number of neighbors of nodes of degree $k$ to be 
quarantined is overestimated).

In order to compare with simulation results performed using the Independent
Cascade Model, we notice that HMF theory provides an epidemic threshold 
which is only approximate for continuous time SIR but is exact 
for the ICM~\cite{PastorSatorras2015}.
Exploiting this correspondence, we assume the validity of Eq.~(\ref{n_Q2})
for ICM dynamics, provided $\beta$  is replaced with the probability $p$
in Eq.~(\ref{tauapp}) for $\tau$.

\bibliography{Quarantine}

\end{document}